%% file: Submission.tex
\documentclass[a4paper,UKenglish,cleveref, autoref, thm-restate]{lipics-v2021}

\usepackage{booktabs}
\usepackage{nameref}
\usepackage{tabularx}
\usepackage[graphicx]{realboxes}
\usepackage{siunitx}
\usepackage{tabularx}
\usepackage{rotating}
\usepackage{cite}
\usepackage{float}

\newcommand{\designiteMetrics}{11}
\newcommand{\understandMetrics}{45}
\newcommand{\jhawkMetrics}{40}
\newcommand{\metricsAfterSC}{63}

\newcommand{\sizeMetrics}{35}
\newcommand{\couplingMetrics}{11}
\newcommand{\cohesionMetrics}{6}
\newcommand{\inheritanceMetrics}{11}

\newcommand{\totalMetricsAnalyzed}{63}
\newcommand{\efaMetricsAnalyzed}{54}
\newcommand{\efaObservationsAnalyzed}{1279}
\newcommand{\efaConstructsExpected}{five}

\newcommand{\totalMetricsResult}{34} 
\newcommand{\totalConstructsResult}{six}

\newcommand{\lowCommunality}{10} 
\newcommand{\wrongFactor}{10} 
\newcommand{\npdMetrics}{9} 

\newcommand{\cfatotalMetricsResult}{24}
\newcommand{\cfaLowLoadings}{10}

\title{Assessing the Construct Validity of Object-Oriented, Class-Level Code Quality Metrics} 

\titlerunning{Assessing the Construct Validity of Code Quality Metrics} 

\author{Hera Arif}{Dalhousie University, Halifax, Canada}{hera.arif@dal.ca}{https://orcid.org/0009-0007-6772-351X}{NSERC Discovery Grant RGPIN-2020-05001.}

\author{Miikka Kuutila}{LUT University, Lahti, Finland}{miikka.kuutila@lut.fi}{https://orcid.org/0000-0002-3695-7280}{}

\author{Paul Ralph}{Dalhousie University, Halifax, Canada}{paulralph@dal.ca}{https://orcid.org/0000-0002-7411-0857}{NSERC Discovery Grant RGPIN-2020-05001.}

\authorrunning{H. Arif, M. Kuutila, and P. Ralph}

\Copyright{Hera Arif, Miikka Kuutila, and Paul Ralph}



\category{Technical Track Paper}

\relatedversion{} 

\supplement{}

\supplementdetails[subcategory={}, cite={}, swhid={}]{Dataset}{https://doi.org/10.5281/zenodo.20278447}

\nolinenumbers 

\EventEditors{Robert Feldt, Maria Paasivaara, Daniel Mendez, Stefan Wagner, and Marvin Mu\~{n}oz Bar\'{o}n}
\EventNoEds{5}
\EventLongTitle{20th International Symposium on Empirical Software Engineering and Measurement (ESEM 2026)}
\EventShortTitle{ESEM 2026}
\EventAcronym{ESEM}
\EventYear{2026}
\EventDate{October 8--9, 2026}
\EventLocation{Munich, Germany}
\EventLogo{}
\SeriesVolume{394}
\ArticleNo{10}

\begin{document}

\maketitle

\begin{abstract}
\emph{Background}: Code quality metrics are intended to measure latent properties of software source code. Although numerous code metrics have been proposed and used, their construct validity is rarely evaluated. Thus, the extent to which code metrics actually measure what they claim to measure is often unclear.

\emph{Aim}: Drawing from modern measurement theory, we investigate the construct validity of common class-level, object-oriented code quality metrics. 

\emph{Method}: As code quality metrics are intended to reflect latent attributes, such as cohesion and coupling, we identified the factor structure of code quality metrics using \textit{Exploratory Factor Analysis} (EFA). The metrics were extracted from the Apache Maven project by three software tools: \textit{Designite, JHawk,} and \textit{Understand}. The factor structure was later verified using \textit{Confirmatory Factor Analysis} (CFA) on 22 randomly selected open source projects meeting a predetermined eligibility criteria.

\emph{Results}: \cfatotalMetricsResult{} code quality metrics that correspond to \totalConstructsResult{} constructs: \textit{Cohesion}, \textit{In-Coupling}, \textit{Out-Coupling}, \textit{Size}, \textit{Sub-Inheritance (related to subclasses)}, and \textit{Sup-Inheritance (related to superclasses)} were revealed in the underlying factor structure. Ten metrics did not correspond to any known dimension of software quality and were removed in the exploratory analysis. Ten additional metrics exhibited low loadings in the confirmatory analysis, suggesting their removal from the final measurement model. \textit{Size, Cohesion, Inheritance,} and \textit{Coupling} were the constructs retained, with subcategories identified for \textit{Inheritance} and \textit{Coupling}.

\emph{Conclusions}: Our results strongly support the construct validity of 24 code quality metrics. 
\textit{Coupling} and \textit{Inheritance} are revealed as multidimensional constructs, since they require measuring two different concepts, revealed as sub-categories in our analysis, and \textit{Complexity} may be better explored in a multilevel model. Our results also corroborate the relationship between \textit{Cohesion}, \textit{Size}, and \textit{Out-Coupling} which can be further explored in a structural model. Some metrics from the Chidamber \& Kemerer metrics suite are found to perhaps be measuring different constructs than intended. 
Our results reveal the need for creating or integrating metrics that reflect existing constructs but measure fundamentally different properties to improve the content validity of the measurement model. Additionally, we provide useful recommendations for researchers, developers, and tool providers which stem from theoretical and empirical justification. Overall, our study demonstrates the value of applying modern measurement theory and latent variable modeling in validating software code quality metrics.

\end{abstract}

\section{Introduction}
\label{sec:introduction}

Code quality metrics provide quantitative measures that evaluate attributes of a software system. Here, code quality refers to the characteristics of source code that influence its overall quality, including complexity, modifiability, understandability, and testability. Accordingly, efforts to improve software quality generally target these quality attributes, either explicitly or implicitly~\cite{Tempero2026}.
Code metrics are used by developers to provide insights about the attributes of software, and are a popular topic of interest for software developers and researchers.  
Despite the proliferation of proposed metrics, metrics research suffers from several persistent challenges. 
The software development industry appears to rely on metrics in a very limited manner~\cite{Tempero2026}; while measurements in other fields form the foundation of a successful system, software developers often favor popular, easy to compute metrics~\cite{Sharma2020Need, Ralph2018Construct}. 
More fundamentally, a key insight in both recent philosophy of science and theory of measurement is that observation is intrinsically problematic~\cite{Ralph2024Metrology}. Most observations are mediated by instruments and measure the effects of ``latent'' structures rather than the structures themselves. In this vernacular, source code is a \textit{latent structure} because we cannot directly observe its execution without instruments (e.g. computer screen, IDE, debugger, terminal window). Important aspects of code such as size, complexity, coupling, and cohesion are \textit{constructs} because, again, we cannot observe them directly. We \textit{operationalize} a construct using one or more \textit{metrics} (e.g. a script that counts lines of code). 

Thinking about code metrics this way highlights several problems: 
\begin{itemize}
    \item The instrument is defective or does not do what the researcher thinks it does (e.g. script for counting lines of Java code that accidentally counts commas instead of semicolons).
    
    \item The instrument measures the property correctly, but the property is not caused by the target construct (e.g. the researcher aims to measure complexity, but counts lines of code, which is caused by the system's size rather than its complexity). 
    
    \item A multidimensional property like coupling (which has direction, strength, and granuality~\cite{Burrows2010}) is operationalized using a single instrument (e.g. Chidamber \& Kemerer's {\sc cbo}~\cite{Chidamber1994Metrics}), which considers incoming and outgoing links as having equivalent strength, disregarding distinctions between these dimensions).
    
    \item An instrument correctly measures the target property, but the property is caused by multiple constructs (e.g. the McCabe~\cite{Mccabe1976Complexity} and Halstead~\cite{Halstead1977} metrics are commonly used to measure complexity, but they are highly correlated with the lines of code in a system~\cite{Graves2000}).
\end{itemize}
Construct validity concerns problems like these--validation refers to the process of ascertaining if an indicator actually measures the aspect it proposes to measure (theoretical validation) and asserting that the indicator relates to other similar measures in an expected way (empirical validation)~\cite{Ralph2018Construct}. 
Assessing how well an indicator operationalizes a construct is crucial for the integrity of a measurement. 
Given the abundance of existing and newly proposed metrics, without a corresponding emphasis on validating them, it becomes challenging to assess how well a metric aligns with the underlying construct it intends to measure. 
Currently, metric's measurement is mostly dependent upon what a software developer understands the metrics definition to be~\cite{Tempero2018Framework, Sharma2018}. For example, metrics with the same name are implemented by two tools differently and differ in the resulting measurement values if applied to the same software code (see \S\ref{ssec:assumptionsfa}). Moreover, code quality metrics defined at the method-level are being used to infer information about class-level attributes--the results of a method level complexity metric ({\sc cc}~\cite{Mccabe1976Complexity}) is being averaged to analyze the complexity of a class as a whole (see \S\ref{ssec:designfa}). This results in a bad metric value being masked and compensated by good metric values, influencing the overall result. Another example is that of Maintainability Index ({\sc mi)} which is considered a proxy of software maintainability, however, a multifaceted concept like maintainability cannot be fully captured by a single metric~\cite{Dag2023}. In fact, metrics chosen to represent maintainability may influence the outcome more than the actual maintainability of code~\cite{Dag2012}. This may mislead professionals by providing them a false sense of confidence in their decisions, or discourage them from relying on code metrics when making important decisions altogether~\cite{Tempero2026, Dag2012, Chowdhury2022}, undermining the very purpose of code metrics, which is to offer scientifically grounded evidence for the measurement of concepts.

Thus, the aim of this study is to investigate the validity of common code quality metrics and the research question of this study is as follows:

\begin{quote}
\textbf{Research Question:}
Do object-oriented, class-level, code quality metrics measure the latent quality attributes they are supposed to measure?
\end{quote}

To accomplish this, we conduct an \textit{Exploratory Factor Analysis} (EFA)--a statistical approach used to discover ``factors'', ``constructs'', or ``latent variables'' from a set of variables (in our case code quality metrics) by analyzing the correlations between them~\cite{Hair2013Multivariate}. This is followed by a \textit{Confirmatory Factor Analysis} (CFA)--which tests the model revealed by the EFA to assess how well the specification of factors and variables represents actual data~\cite{Hair2018Multivariate}.

\section{Related Work} 
\label{sec:relatedwork}
In this section, we provide an overview on the theory of measurement, the current state of metrics literature and prior research related to our study. 

\subsection{Contemporary Theory of Measurement} \label{ssec:measurementTheory}

Measurement is the principled assignment of numbers to reality~\cite{Campbell2013Physics}. A metric is the process of representing a phenomenon numerically. Operationalizing unobservable phenomena using a single metric is philosophically and methodologically outdated because it assumes measurement is unproblematic and ignores mono-method bias~\cite{Ralph2024Metrology}. Instead, contemporary measurement theory suggests constructing measurement models that operationalize each phenomenon (latent variable) as the shared variance of multiple diverse metrics. 

Software maintainability, flexibility, understandability are external quality dimensions that researchers acknowledge cannot be observed directly. Whereas, software complexity, coupling, cohesion, are specific internal quality criteria that researchers strive to measure directly using new or improved metrics. However, multiple metrics differ slightly and sometimes significantly when measuring the same construct (see \S\ref{ssec:assumptionsfa}). This suggests that metrics are indirect measures and should not be used as a direct representation of the construct. Thus, code quality is a \textit{latent structure} and metrics can be considered indicators that measure a theoretical concept that cannot be measured directly~\cite{henseler2020composite}.

\subsection{Current State of Code Quality Metrics Research}
A code quality metric, in essence, measures attributes of a software by analyzing its source code. 
For example, {\sc loc} measures the lines of code of software code.
Initially, research related to code quality metrics focused on size and complexity metrics like {\sc loc}, {\sc cc}~\cite{Mccabe1976Complexity}, and the Halstead metrics~\cite{Halstead1977} but then shifted to object-oriented (OO) metrics like the C\&K~\cite{Chidamber1994Metrics}, MOOD~\cite{Abreu1994Object},
and Martin's design metrics~\cite{Martin1995Designing, Sharma2020Need}. 
Researchers are increasingly trying to determine factors that are holding back metrics research and how these problems can be fixed~\cite{Tempero2026}. Some common criticisms are as follows.
\textbf{Limited use:} Code quality metrics' impact on software development is underexplored, with metrics like {\sc loc} and {\sc cc} used commonly; while others ignored or used rarely~\cite{Sharma2020Need}.
\textbf{Lack of precise definitions and guidelines:} Every new metric is expected to provide a sound theoretical foundation and a precise definition, which seem to be lacking in the present literature~\cite{Sharma2020Need}.
Additionally, there are numerous discrepancies in the naming conventions of metrics. For example, Depth of Inheritance Tree ({\sc dit}) is also called Depth of Inheritance ({\sc dih}) but both these metrics measure the maximum length of a tree from node to root~\cite{Saraiva2012Aspect}. 
\textbf{Shift towards code smells:} More recently, code smells are used as a metaphor to depict patterns commonly linked with poor design and programming practices~\cite{Sharma2018}. However, the idea of code smells departs significantly from contemporary theories of measurement. Code quality is a latent structure whose measurement requires a statistical measurement model that specifies the dimensions of code quality and operationalizes each of them as the shared variance of multiple indicators. The presence or frequency of one or more code smells could constitute some of the indicators in this measurement model. However, focusing only on code smells to measure code quality is insufficient and doesn't exempt the researcher from constructing a sound measurement model and quantitatively assessing its validity.
\textbf{Rise of agentic code:} As software developers incorporate more agentic code generation, a simultaneous increase in assessing the quality of code generated by AI is observed. Recent studies have employed the use of traditional code quality metrics like {\sc mi}, {\sc loc}, and {\sc cc} to understand the quality of agentic code developed by various generative AIs like GitHub Copilot~\footnote{\url{https://github.com/features/copilot}}, ChatGPT~\footnote{\url{https://chatgpt.com/}}, etc.~\cite{Idrisov2024, Nguyen2022}. These studies highlight generative AI's focus on accelerating programming, sometimes even at the expense of code quality.
\textbf{Poor validation:} Over the years, researchers have proposed numerous approaches for validating software metrics. Some researchers propose a set of ``validity criteria'' to assess metrics~\cite{Weyuker1988Evaluating}. Weyuker et al.~\cite{Weyuker1988Evaluating} define a set of ``properties'' for complexity metrics; and while Cherniavsky et al.~\cite{Cherniavsky1994On} and Fenton~\cite{Fenton1994Software} disagree with this approach, Chidamber et al.~\cite{Chidamber1994Metrics} support it.
Another example is the findings of a  literature review by Meneenly et al.~\cite{Meenly2013Validating} that reports 47 validation criteria for software code quality metrics highlighting the ad-hoc nature of metrics validation as well as conflicting and subjective perspectives among researchers. Some studies interpret correlation as validation, considering metrics valid if they are correlated with other metrics measuring the same construct~\cite{Fenton1994Software}; however, correlation and validation are not interchangeable. 
Some empirical studies consider metrics as predictors of external features and consider this relationship as evidence of metric validity. Basili et al. ~\cite{basili1996Validation} investigated the C\&K metrics suite's~\cite{Chidamber1991Towards} effectiveness as predictors of fault proneness and Briand et al. ~\cite{briand2000} investigated the relationship between object-oriented metrics and the likelihood of fault detection in classes. They found that metrics were effective~\cite{basili1996Validation} but redundant~\cite{briand2000} as they captured similar attributes of the software. 
Emam et al.~\cite{emam2001} and Gil et al.~\cite{Gil2017} studied the relationship between external quality attributes and size metrics. They recommended that future studies take the confounding effect of size into consideration when investigating the validity of metrics.
Researchers have also attempted to create unified frameworks for software quality measurement (such as ~\cite{briand1996property, briand1997cohesion, Briand1999Unified, Kitchenham1995Towards}). These studies emphasized the need for stronger theoretical foundations for software measurement and the need for further research into scientifically sound methods for measuring software to ensure valid measurements.
They also highlighted that metric's validation remains an unresolved challenge that must be addressed to achieve reliable measurement of software quality.

The underlying \textit{factor structure}--the relationships that exist between variables (code quality metrics) and the constructs they are supposed to be measuring, is not taken into consideration in most metrics research and unvalidated metrics are used as \textit{direct measures} to measure code quality constructs (such as size, complexity, and cohesion)~\cite{Ralph2018Construct}.
The metrics literature appears to overlook the distinction between direct measures and indirect proxy measures and largely focuses on identifying a single direct metric that can be applied to every objective~\cite{Fenton1994Software, Bouwers2012}. Consider two variants of C\&K's {\sc lcom}~\cite{Chidamber1994Metrics}: {\sc lcom2}~\cite{Chidamber1991Towards} and {\sc lcom5}~\cite{Henderson1996Coupling}; they are both meant to measure the cohesion in software code, so a strong correlation between them is expected. In fact, this is not the case (see \textit{\nameref{sec:discussion}}). These issues are addressed by ``construct validity''--the degree to which a warranted inference can be made for a construct using a measure. Construct validity includes assessing: \textit{face validity}--does a measure ``on its face'' seem to be a good measure of the construct?, \textit{content validity}--is the measure representative of relevant domain related to the construct?, \textit{convergent validity}--how similar is the measure with other measures it should be theoretically similar to?, \textit{discriminant validity}--how different is the measure with other measures it should theoretically be different to?, and \textit{predictive validity}--does the measure predict something it should theoretically be able to predict?~\cite{trochim2007}. 

\subsection{Prior Research on Factor Analysis and Code Metrics} 

Munson et al.~\cite{Munson1992} performed an EFA to reduce a number of complexity metrics into a smaller set of independent complexity domains. They analyzed 106 modules and 16 software complexity code metrics produced by their Ada Metric Analyzer to group complexity into five constructs: control, size, information content, modularity, and data structure. Al et al.~\cite{Al1994} conducted a study to identify the efficiency of traditional metrics in measuring the complexity of OO code. 
They used the CPPOOM (C++ ObjectOriented Metric) tool to calculate metrics on four C++ systems, analyzing 43 trees and 222 classes, to reveal four complexity constructs. They concluded that traditional metrics loaded on a single factor and were correlated with other traditional metrics and encouraged further research on the topic. Cook et al.~\cite{cook1994} conducted a study to analyze ten releases of a real-time telephone switching system written in a macro-assembly language. They calculated 18 complexity metrics and conducted a principal components analysis to reduce dimensionality. The results revealed four complexity domains: size, information flow into/out of functions, and control flow. 

Prior research in this domain seems to focus on a single construct of complexity with the aim of finding sub-categories or performing dimension reduction ~\cite{Munson1992, Al1994, cook1994}. These studies are conducted over 30 years ago with a small dataset of code quality metrics and lack implementation details. 
We aim to build upon this research by considering numerous metrics and constructs, and providing a publicly accessible dataset with a comprehensive replication package that can be used to re-implement our analysis and ensure accuracy and reliability. 

\section{Methodology} 
\label{sec:methodology}

Assessing the construct validity of code quality metrics requires a (rudimentary, incomplete) theory of code quality that can be generated inductively from data by examining the factor structure of metrics using factor analysis. Briefly, our methodology involves selecting some source code to be fed into software tools that calculate code quality metrics. These metrics are screened and excluded if not appropriate for a factor analysis. The construct validity of the resulting metrics is then assessed iteratively using an exploratory and a confirmatory factor analysis revealing six constructs comprising 24 metrics.
All statistical analysis is performed with R version 4.3.2, using the R packages \texttt{psych}~\cite{psych} and \texttt{lavaan}~\cite{lavaan12, lavaan25}. 

\subsection{Exploratory Factor Analysis}
\label{ssec:efamethod}
As the name suggests, an EFA ``explores'' the underlying factor structure represented by variables to develop a measurement model~\cite{Hair2018Multivariate}. Our objective is to assess the construct validity of common, object-oriented, class-level, code quality metrics. We followed the method outlined by Hair et al.~\cite{Hair2013Multivariate} to conduct an EFA by performing the following steps. 

\subsubsection{Designing the Factor Analysis}
\label{ssec:designfa}

To assess the factor structure of the code quality metrics, we must compute metrics for some code. Representative sampling is neither practical nor desirable for the initial, exploratory stage (see \textit{\nameref{ssec:limitations}}). Rather, we want a single project that is large enough for the factor structure to emerge but small enough that the codebase is comprehensible. To avoid becoming overwhelmed by complexity, we begin by focusing on class-level Java metrics because both the language and quality concepts would be familiar to many researchers interested in software metrics. This includes many popular metrics (e.g. C\&K~\cite{Chidamber1994Metrics}, L\&K~\cite{Lorenz1994Object}, MOOD metrics~\cite{Harrison1998Evaluation}).
\begin{itemize}
    \item \textbf{Tools Selection Criteria:} The software tools must calculate a wide range of class-level metrics, and must be able to analyze large projects containing mainly Java code.
    \item \textbf{Project Selection Criteria:} In a preliminary, exploratory phase, we only require a project with enough observations for the factor structure to emerge--which is a property of the variables themselves and not the objects of the study (see \textit{\nameref{ssec:limitations}}). Additionally, the project should contain at least 10 observations per variable~\cite{Hair2013Multivariate}.
    \item \textbf{Class Selection Criteria:} Exclude any classes for which one or more selected tools cannot calculate \textit{any} metrics.
    \item \textbf{Metrics Selection Criteria:} Metrics should be numerical, at an ordinal, interval, or ratio level. They should measure something about the code structure at class-level that corresponds to an identifiable construct, and should not be a function of method-level metrics.
\end{itemize}

The software tools selected were: Designite,\footnote{DesigniteJava Enterprise (v. 2.1.2), \url{https://www.designite-tools.com/}} Understand,\footnote{Understand (v. 6.1), \url{https://www.scitools.com/}} and JHawk\footnote{JHawk (v. 6.1.7), \url{http://www.virtualmachinery.com/jhawkprod.htm}} which calculate \designiteMetrics{}, \understandMetrics{}, and \jhawkMetrics{} class-level code quality metrics respectively, yielding a total of 96 metrics. These tools are popular for metrics research because they each contain a wide range of metrics found in the literature and are calculated at different levels (e.g. file, class, method level). There could be other tools that meet our selection criteria. It is not necessary for our purposes to include every tool that meets the above criteria.
We selected the \textit{Apache Maven} project,\footnote{\label{footnote:maven}Apache Maven (v. 2.0), \url{https://github.com/apache/maven}} a popular open-source Java tool used to manage external dependencies and automatically build software projects as the selected project should contain at least 960 classes ($10*96$).  
The Apache Maven source code was downloaded from its GitHub repository~\textsuperscript{\ref{footnote:maven}} and compiled. It contains around 1,300 classes which are sufficient for our analysis. 
Special Java classes like enums, anonymous, or nested inner classes are not recognized by some tools so they are excluded from our analysis (e.g. Understand does not calculate {\sc lcom} metrics for enums).
Non-numerical metrics such as {\sc name of super class} were excluded. Some metrics were calculated at the method-level and summed to produce a class-level measurement. For example, {\sc cc} is calculated at the method-level to estimate class complexity. We exclude such metrics because there are more sophisticated ways of handling this situation (see \textit{\nameref{ssec:futurework}}).
To determine which construct each metric ostensibly reflects, we refer to the documentation provided by the software metric tools.
For metrics that were not categorized into any constructs, we reviewed the metrics literature. A complete list of the included and excluded metrics and their mapped constructs is provided in the replication package (see \textit{\nameref{sec:dataavailability}}). 
Based on the selection criteria, 33 metrics out of the 96 metrics calculated were excluded. 
The final dataset contains measurements from \sizeMetrics{} size metrics, \cohesionMetrics{} cohesion metrics, \inheritanceMetrics{} inheritance metrics, and \couplingMetrics{} coupling metrics--\metricsAfterSC{} in total. We aim to classify these metrics into \efaConstructsExpected{} factors: size, cohesion, inheritance, in-coupling, and out-coupling, following the classification provided by the software tools. Table \ref{tab:factorsDefinition} shows a brief explanation of these factors in terms of class-level.

\input{Tables/5factors}

\subsubsection{Test Assumptions of Factor Analysis}
\label{ssec:assumptionsfa}

Factor analysis conceptually assumes that an underlying factor structure exists in the data that is to be analyzed.
The purpose of this study is also to reveal this underlying factor structure and to assess its validity. Factor analysis  assumes a homogeneous sample of measurements which our dataset complies to.
The correlation matrix for our dataset has numerous correlations greater than 0.3 indicating a correlated set of variables appropriate for factor analysis~\cite{Hair2013Multivariate}.
Since the tools calculate some of the same metrics, some metrics (e.g. \textit{Size.NOM.Understand} and \textit{Size.NOM.JHawk}) have a perfect linear relationship (singularity) indicating a valid measurement. More surprisingly, seemingly different metrics are identical, e.g. \textit{Inheritance.DIT.JHawk} and \textit{Inheritance.CountSup.JHawk} have a perfect linear relationship and \textit{Size.CountDeclMethodDefault.Understand} and \textit{Size.CountComment.JHawk} are highly correlated (\textit{r} = 0.83). Nine such metrics were removed as they caused a ``not positive definite'' (NPD) which is not factorable.  Excluding these nine metrics has little effect because they are so tightly correlated with remaining metrics. The Kaiser–Meyer–Olkin (KMO) test to assess sampling adequacy found the dataset ``meritorious''~\cite{Kaiser1974Little} ($KMO = 0.84$). Bartlett's test of sphericity was significant ($p<0.001$), indicating correlated variables. The final dataset consists of \efaObservationsAnalyzed{} observations (classes) and \efaMetricsAnalyzed{} variables (code quality metrics)--a ratio of approximately 20 observations per variable, which is higher than the recommended minimum sample size and thus acceptable for conducting factor analysis~\cite{Hair2013Multivariate}.

\subsubsection{Derive Factors and Assess Fit}
\label{ssec:fitfa}

We used multiple methods to determine the number of factors to extract:
parallel analysis~\cite{Horn1965Rationale} suggests 22 factors, the scree plot~\cite{Cattell1966Scree} suggests 2 or 8 factors, and the Kaiser criteria~\cite{Kaiser1960Application} suggests 8 factors. These methods could overestimate the number of factors (an over-fit model) in very large datasets like ours and involve some degree of subjectivity~\cite{Cattell1966Scree}. 
A combination of methods can be used to make a definitive decision and prior theoretical knowledge of the expected number of factors can be very beneficial in making this decision~\cite{Costello2005Best}. 
According to theory, we predict five factors: size, cohesion, inheritance, in-coupling, and out-coupling (Table \ref{tab:factorsDefinition}). So, we begin with eight factors since that is the largest reasonable estimate of the number of factors (the parallel analysis estimate of 22 factors is prima facie unreasonable). The number of factors retained should have an explained variance of at least 60\%~\cite{Hair2013Multivariate}. Our initial eight-factor model explained 71\% of the variance.

\subsubsection{Interpret Factors}
\label{ssec:interpretfactors}

We rotate the solution using an \textit{oblique}, ``oblimin'' rotation as it assumes factors are correlated~\cite{Hair2013Multivariate}.
The construct validity of the metrics is assessed iteratively using an exploratory factor analysis. The factor solution obtained needs refining as it is overcomplicated and contains numerous unacceptable factor loadings--a \textit{factor loading} explains how much of a variable's variance can be explained by the factor~\cite{Hair2013Multivariate}. A high loading means that the variance explained for a variable is sufficient for it to have a considerable relationship with the factor. The factor solution also contains low communalities (h2) which is the amount of variance in a variable that can be explained by the factor solution extracted by the EFA~\cite{Field2017Discovering}.
We perform the following steps to refine the model.
\begin{enumerate}
    \item \textbf{Remove variables with low loadings ($<$ 0.5) and low communalities ($<$ 0.5) in the eight-factor model:} We iteratively removed such variables, rerunning the factor analysis after each removal until no such variables remain (the loadings and communalities change every time a factor analysis is run). Eight variables were removed, increasing the variance explained by the factor model to 83\%.
    
    \item \textbf{Reduce from eight to six factors:} In the eight-factor model, the \textit{Inheritance} and \textit{Size} variables each loaded on more than two factors, suggesting an over-fit, so we re-ran the factor analysis with fewer factors to investigate. Despite some cross-loadings, we could label the factors more easily and clearly in the six-factor model, which map well into our a-priori constructs: \textit{Size, Cohesion, In-Coupling, Out-Coupling}, and two types of inheritance--\textit{Sub-Inheritance} and \textit{Sup-Inheritance} that measure the characteristics of a class related to its sub-classes and super-classes respectively. The variance explained by the factor model decreased to 74\%.\footnote{The variance explained by a factor analysis increases monotonically with the number of factors extracted, so reducing the number of factors will \textit{not} increase the variance explained even if the model is better in other ways.}

    \item \textbf{Remove variables with low loadings ($<$ 0.5) and low communalities ($<$ 0.5) in the six-factor model:} Similar to before, we iteratively remove such variables, re-running the factor analysis after every removal. Two variables were removed, increasing the variance explained by the factor model to 76\%.

    \item \textbf{Remove variables loading on the incorrect factor:} We analyzed the model for variables loading on the wrong construct, deleting them in order of lowest communalities, and ignoring cross-loadings hoping they would integrate into a single high loading. Ten metrics were removed, increasing the variance explained by the model to 85\%.

\end{enumerate}

\subsubsection{Exploratory Factor Analysis Results}

Table \ref{tab:EFAFinal} shows the final solution, which  explains 85\% of the total variance with the six factors extracted, which is above the recommended minimum threshold~\cite{Hair2013Multivariate}. Each factor contains at least three variables, and correspond to the theorized classification (Table~\ref{tab:factorsDefinition}), with sub-categories revealed for \textit{Inheritance} i.e. \textit{Sub-Inheritance} (related to sub-classes), and \textit{Sup-Inheritance} (related to super-classes). The Cronbach's $\alpha$ for the overall model is $\alpha=0.92$, which is considered ``excellent''. The Cronbach's $\alpha$ for each individual factor in the model also has acceptable values indicating internally consistent variables in the model~\cite{Cronbach1951Coefficient}. 

\input{Tables/EFAFinal}

\subsection{Confirmatory Factor Analysis and Results} 
\label{sec:cfaMethod}

We followed the method outlined by Hair et al.~\cite{Hair2018Multivariate} to test the pre-specified measurement theory obtained from the EFA and assess how well the theory fits population data. The steps followed are described below.

\subsubsection{Define Individual Constructs} 
\label{ssec:defIndCons}

Considering the \totalMetricsResult{} code quality metrics measuring six constructs revealed as the result of the EFA, the conceptual definitions of constructs match well with the item wordings, and the relationships are corroborated by previous literature, scholarly peer-review, and tool documentation, thus establishing reasonable face validity. Some metrics seem to clearly be measuring a specific construct (for example, {\sc count line} measures \textit{Size}); and some metrics seem to be conflated (for example, {\sc cbo} measures both incoming and outgoing dependencies and seems to be conflated between \textit{In-Coupling} and \textit{Out-Coupling})--therein lying the motivation of this paper. 

\subsubsection{Develop the Overall Measurement Model}

The six latent constructs may correlate with each other as they would in real software where quality dimensions are congeneric. An indicator variable loads on a single construct and error variances between indicators exist but are not allowed to correlate.
Every construct is identified with a minimum of three indicators per construct asserting an ideal ``over-identified model''--a model with more degrees of freedom than paths to be estimated ($dof=512$). In the proposed model, all constructs are hypothesized as \textit{reflective} making the direction of causality from the latent construct to the measured indicator variables.

\subsubsection{Design a Study To Produce Empirical Results}

We used the SEART GitHub Search Engine\footnote{\url{https://seart-ghs.si.usi.ch/}} to collect a sample of Java projects by applying minimum selection thresholds. These criteria were intended to identify professionally developed and actively maintained projects, rather than hobby or dummy projects.
We obtained 71 Java projects, out of which, 64 projects could be compiled using the popular build tools Maven,\footnote{\url{https://maven.apache.org/}} Gradle,\footnote{\url{https://gradle.org/}} or Ant\footnote{\url{https://ant.apache.org/}}. Any project used in the exploratory analysis cannot be used in the confirmatory phase; thus, one project was excluded. Since we need compiled code to conduct the analysis, we selected the 63 projects and shuffled them using Python's \texttt{random.shuffle} method, and downloaded and compiled them. We need a minimum of 340 classes to conduct the CFA ($\totalMetricsResult{} * 10$)~\cite{Hair2018Multivariate}. Since we can get a much larger dataset, we decided to use workload as the deciding factor of sample size. We aim to complete data collection in two weeks (80 hours). Downloading, compiling, and preparing each project takes approximately 3.75 hours.

\begin{equation}\label{eq1}
Number of projects = \frac{80 hours}{3.75 hours/project} = 21.33
\end{equation}

Thus, 22 projects (\ref{eq1}) were included in our analysis.~\footnote{All the data collection was completed by May, 2025} The smallest project included had 506 classes, and the largest project included had 19,833 classes ($M=3063$).
A list of the parameters set and additional details about the included projects can be found in the replication package (See \textit{\nameref{sec:dataavailability}}).
A process similar to the one described in \S\ref{ssec:designfa} was followed to extract the code quality metrics collected by Designite, JHawk, and Understand and combine them into a single spreadsheet. Creating a combined dataset does not affect the results of the analysis since we are performing instrumentation not prediction. Given the number of indicators and constructs, our sample size of approximately 100,000 is more than sufficient for analysis. 

\subsubsection{Assess Measurement Model Validity}

A confirmatory factor analysis is conducted on the dataset and a six construct measurement model is produced (Table \ref{tab:loadings}). The purpose of the factor analysis performed in this study was primarily to evaluate the relationship between commonly used software code metrics and their hypothesized constructs, and not to optimize a measurement instrument with a maximized global fit. Thus, some of the metrics showing weaker loadings were intentionally retained during model evaluation to assess their behavior within the proposed construct structure. Consequently, we also do not report overall fit indices, as we are investigating whether metrics measure the latent quality attributes they are supposed to measure, and not what is the most accurate way to do so. 

\input{Tables/loadings}

We now assess the validity of the measurement model obtained by assessing the following.

\begin{enumerate}
    \item \textbf{Face, Nomological, and Predictive Validity:} Face validity was established before the factor analysis was conducted (see \S\ref{ssec:defIndCons}). Nomological validity can be estimated by looking at the correlation between constructs and concluding if they make sense. All correlations between the constructs in the model are statistically significant at $p<0.05$, as expected. There is a moderate positive correlation between \textit{Out-Coupling} with \textit{Cohesion} ($r=0.33$) and \textit{Size} ($r=0.31$). Other correlations are positive and low. This also satisfies predictive validity in a way that constructs posit predicted causal relationships.

    \item \textbf{Content Validity:} We are limited to the metrics provided by Designite, JHawk, and Understand. These tools are frequently used in research related to software code quality, so we presume that these metrics cover the breadth of the constructs.

    \item \textbf{Convergent Validity:} Convergent validity is estimated using several techniques (Table \ref{tab:loadings}). The \textit{standardized factor loadings} are acceptable for most variables ($>0.7$). Variables with low loadings load majorly on \textit{Size} suggesting further investigation into their inclusion in the model. Two loadings are greater than one (have an error variance less than zero)--\textit{Cohesion.LCOM.Understand} ($e=-0.004$) and \textit{Sup-Inheritance.CountSup.JHawk} ($e=-0.007$), a ``heywood case''. This can occur as a compensatory mechanism for high residual correlation between highly correlated variables that causes an inflated loading~\cite{Brown2015, Joreskog1994}.  
    The \textit{AVE} and \textit{CR} of all constructs in the model have acceptable values--($>0.5$) and ($>0.7$) respectively, indicating good convergence and internally consistent variables in constructs~\cite{Hair2018Multivariate}. Overall, problematic variables do not appear to be significantly harming model fit or internal consistency and sufficient evidence of convergent validity is provided.
    
    \item \textbf{Discriminant Validity:} We compared the squared correlations between constructs with the AVE values for all construct pairs in the model and found that the AVEs were greater, indicating good discriminant validity (Table \ref{tab:loadings}).

\end{enumerate}

\section{Results}
\label{sec:results}

In this section we answer our research question directly:
\label{results}
\begin{quote}
\textbf{Research Question:}
Do object-oriented, class-level, code quality metrics measure the latent quality attributes they are supposed to measure?

\textbf{Answer:}
At least \cfatotalMetricsResult{} object-oriented, class-level metrics appear to measure the construct they are intended to measure, while the inclusion of \cfaLowLoadings{} metrics needs further investigation.
\end{quote}

\begin{figure}[H]
\centering
\centerline{\includegraphics[scale=0.15]{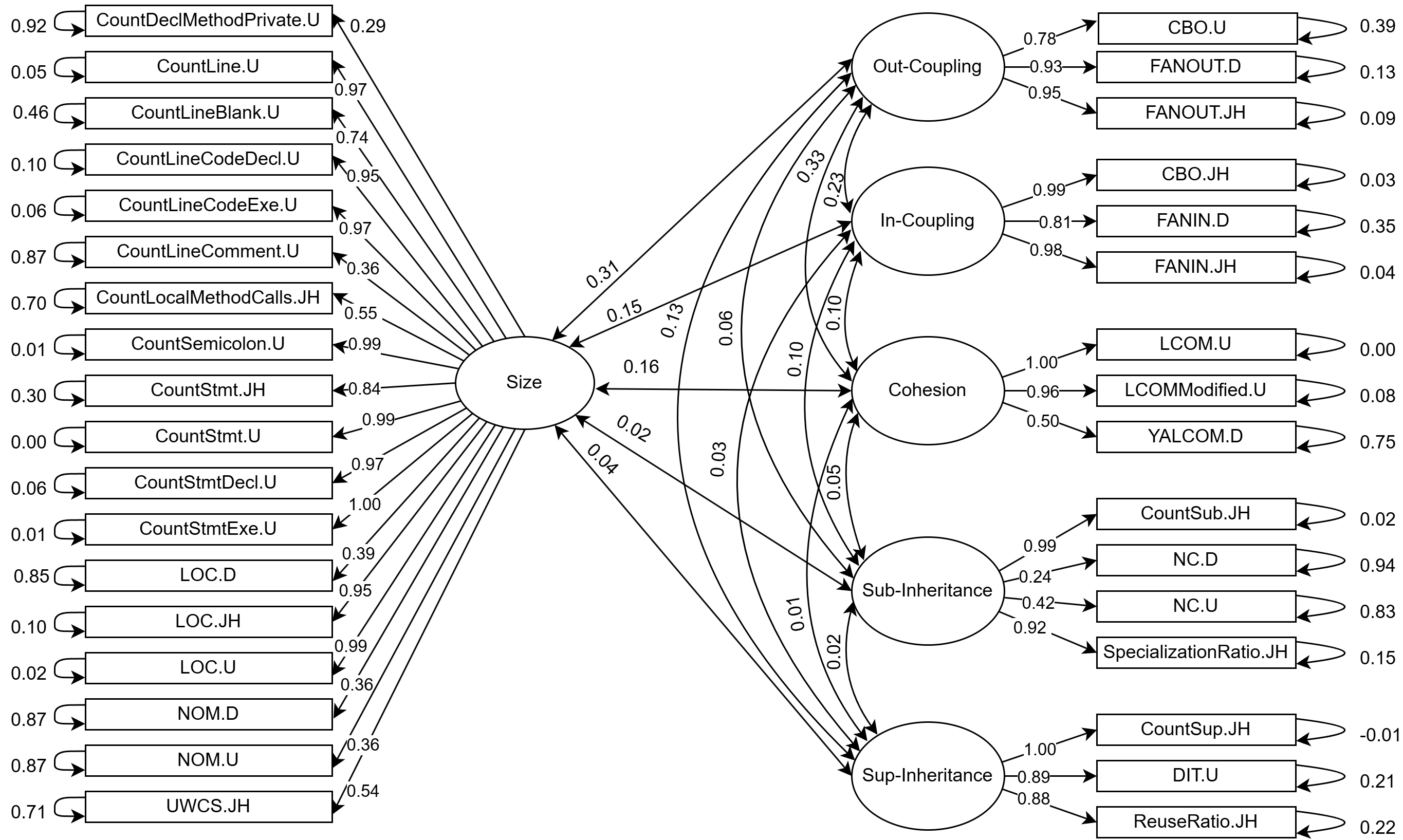}}
\caption*{\scriptsize Note: The acronym at the end of the metrics corresponds to the tool being used, with U corresponding to Understand, JH to JHawk and D to Designite}
\caption{Measurement Theory Model for Code Quality Metrics}
\label{fig:mtm}
\end{figure}

Figure \ref{fig:mtm} illustrates the measurement model with six latent constructs measured reflectively by \totalMetricsResult{} indicator variables.
To be more precise, of the \totalMetricsAnalyzed{} metrics we assessed:
\begin{itemize}
    \item \cfatotalMetricsResult{} metrics seem to reflect the construct they are intended to reflect; i.e. they measure what they are supposed to measure.
    \item \lowCommunality{} metrics do not reflect any of the constructs in the model (including the construct they are intended to measure); i.e. they do \textit{not} measure what they are supposed to measure.  
    \item \wrongFactor{} metrics loaded on the wrong factor; i.e. they do \textit{not} measure what they are supposed to measure.  
    \item \npdMetrics{} metrics caused an NPD matrix; i.e. our analysis is inconclusive regarding whether they measure what they are supposed to measure.
    \item \cfaLowLoadings{} metrics load low in the CFA, i.e. their inclusion in the model needs further investigation.
\end{itemize}

\section{Discussion} 
\label{sec:discussion}

Our results are surprising in several ways. First, numerous widely cited code quality metrics appear not to measure the construct they purport to measure. For example, multiple metrics proposed by Chidamber et al.~\cite{Chidamber1991Towards} were excluded from the final measurement model--\textit{Coupling.RFC.Understand} was removed from the EFA because of low communality and \textit{Cohesion.LCOM2.JHawk} was removed from the EFA because it loaded on \textit{Size}. Additionally, \textit{Sub-Inheritance.NC.Understand} and \textit{Sub-Inheritance.NC.JHawk} loaded low in the CFA suggesting their removal. 
The fact that so many metrics could be proposed, implemented, and used without anyone demonstrating that they reflect the underlying construct of interest suggests a worrisome lack of concern for construct validity in code metrics research. Researchers should shift their perspective of \textit{proposing} software metrics as the ultimate research objective to instead focusing on \textit{validating} software metrics by adhering to the science of the theory of measurement to ensure that reliable and valid measurements are produced~\cite{Graziotin2022Psychometrics, Tempero2026, Ralph2024Metrology}. 

Second, the \textit{lack} of multicollinearity problems is surprising. Highly (or perfectly) correlated variables prevents the factor structure from computing (the dreaded ``NPD matrix''). Meanwhile, the three tools we studied compute many of the same metrics; for instance, all three tools compute {\sc loc}, {\sc nc}, and {\sc nom}. Retaining two implementations of the same metric \textit{should} make the \textit{factor analysis} fail, and in some cases it did, but in many cases it did not. The fact that the model includes, for example, numerous {\sc loc} metrics, suggests non-trivial differences in the way the three tools compute {\sc loc}. As another example, \textit{Inheritance.DIT.Designite} is not highly correlated  with \textit{Inheritance.DIT.JHawk} ($r=0.271$) or  \textit{Inheritance.DIT.Understand} ($r=0.332$). Investigating these differences is beyond the scope of the present study, but something seems wrong here. We retained seven metrics that were calculated by at least two tools: {\sc loc, nc, count stmt, nom, fanin, fanout,} and {\sc cbo}.

Third, the relationships between the code quality constructs identified in our analysis support our theory that the constructs are correlated with each other, but not to the extent that they measure the same underlying concept. We found a moderately positive correlation between \textit{Out-Coupling} with \textit{Cohesion} ($r=0.33$) and \textit{Size} ($r=0.31$). These relationships are also observed in the literature~\cite{Ralph2018Construct} motivating further research, particularly when extending the measurement model to include inter-construct relationships to form a structural model.

Fourth, some metrics are inherently conflated but are popularly used to measure a singular concept. We have established from our analysis that \textit{Coupling} is a multidimensional concept and popular measures (like \textit{Out-Coupling.CBO.Understand}) are conceptually conflated as they measure the incoming and outgoing connections between code entities. Thus, {\sc cbo} measures two different concepts and cannot be combined in an attempt to capture both aspects at once~\cite{Ralph2018Construct}. Our previous discussion (see \S\ref{ssec:measurementTheory}) on code quality metrics becoming increasingly detached from science and theory of measurement seems applicable here: measurements should have clear objectives and specify the entities of interest and their significant attributes to ensure reliable and valid measurements~\cite{Fenton1994Software, Graziotin2022Psychometrics, Bouwers2012}. In the case of software code quality, perhaps the high interrelation of code quality constructs often makes it difficult to determine which indicators correspond to which constructs.  
This observation has also been highlighted by researchers in the past who notice a focus on empirical validation over theoretical validation: 
\begin{itemize}
    \item Fregnan et al.~\cite{Fregnan2019Survey} and Gomez et al.~\cite{Gomez2008} discovered that most metrics research primarily focused on empirical validation through comparisons with existing indicators or by assessing correlations; disregarding theoretical validation. 
    \item Kitchenham et al.~\cite{Kitchenham2010Whats} highlighted that researchers repeatedly empirically validate popular metrics proven to be theoretically invalid; rendering their analysis meaningless.
    \item Stevanetic et al.~\cite{Stevanetic2015} noted that validating a metric’s relevance requires a theoretical justification supported by empirical evidence, which is reportedly missing in current metrics research. 
    \item Briand et al.~\cite{briand1996property, briand2000} emphasized the need for stronger theoretical foundations in software measurement to ensure that metrics accurately measure their intended constructs.
\end{itemize}
Thus, theoretical \textit{and} empirical validation of the relationship between metrics and constructs using appropriate techniques such as factor analysis is both essential and valuable, and should be encouraged~\cite{Tempero2026}.

Fifth, as AI agents become more widely used and AI generated code grows increasingly sophisticated, the need for humans to evaluate, assess, and understand this code also increases. This highlights the need for code that is understandable and readable, with code architecture that is easy for humans to interpret. 
Researchers are using code quality metrics as one way to assess the quality of AI generated code. For example, Murthy et al.~\cite{murthy} analyze the understandability, maintainability, and complexity of AI generated code using {\sc cc}~\cite{Mccabe1976Complexity}. In such studies, metrics are treated as dependent or outcome variables and their effect on a code quality attribute is estimated / predicted. A measurement is only meaningful if it accurately reflects what it is meant to measure. If the measurement lacks validity, the conclusions might be misleading. This highlights the importance of developing valid metrics, benchmarks, and measurement models to support human effort in understanding code quality, which is an objective our research aims to address.

\subsection{Recommendations}
\label{ssec:recommendations}

In light of the findings of this study and an understanding of the theory of measurement which emphasizes that the operationalization of a construct must stem from a theoretical and empirical justification, we offer the following recommendations for researchers, developers, and tool providers.
\begin{enumerate}
    \item \textbf{DO} measure \textit{Cohesion, In-Coupling, Out-Coupling, Size, Sup-Inheritance} and \textit{Sub-Inheritance} using the corresponding metrics shown in Table \ref{tab:loadings}. Both professionals and scientists should prefer these metrics.
    \item \textbf{DO NOT} use the metrics recommended for exclusion from the model due to low / incorrect loadings or low communality; or use them with discretion.
    \item \textbf{DO NOT} use new, unvalidated metrics; rather encourage validation of newly proposed metrics.
    \item \textbf{DO} Estimate constructs using factor scores 
    from a confirmatory factor analysis or similar technique.
    \item \textbf{DO NOT} use individual metrics to estimate constructs.
    \item \textbf{DO NOT} estimate constructs by averaging / aggregating metrics because metrics vary in distribution and relative importance to their corresponding construct. Use factor scores.
    \item \textbf{DO} ask providers of software metrics tools to present factor scores and \textbf{DO} present factor scores in any tools you build.
    \item \textbf{DO NOT} propose new metrics without quantitatively assessing construct and measurement validity using factor analysis or a similar technique. 
    \item \textbf{DO NOT} accept for publication manuscripts proposing more unvalidated metrics, or metric proposals without permanently archived source code (because written descriptions of metric algorithms are often incomplete).
    \item \textbf{DO} use evidence standards (e.g.~\cite{ralph2020empirical}) to design studies and assess manuscripts; \textbf{DO} ensure evidence standards require construct and measurement validity assessment for new metrics. 
\end{enumerate}

\subsection{Implications}

This study serves as a foundational step in the largely under-explored field of assessing the construct validity of software code quality metrics and paves the way for new research opportunities. This study serves as a guiding framework and encouragement for researchers to replicate, generalize, challenge, and extend our findings. To facilitate this, all data used in this study is openly accessible for replication and additional analysis (see \textit{\nameref{sec:dataavailability}}). Researchers may also adopt this approach when developing new metrics to provide evidence of construct validity.
The findings of this study encourage further exploration into the relationships between various software code quality constructs in addition to the relationships between variables measuring these constructs.
Certain aspects of code quality are inherently subjective yet crucial for developers working with code written by others. For instance, Mannan et al.~\cite{mannan2018} reported the need for better measurement and modeling of readability as perceived readability is difficult to quantify and integrate into a measurement model. Therefore, further research is needed to evaluate such attributes and their potential incorporation into a structural model.

Software developers and professionals may prefer qualitative assessments such as experience, expert judgment, intuition, or code smells, over quantitative measures like code quality metrics because of the uncertainty in metrics' implementations and measurements. For example, our results find that C\&K metrics like {\sc rfc, nc, lcom2}~\cite{Chidamber1991Towards} may not accurately capture the quality aspects they are intended to measure. 
On the other hand, developers and professionals are more likely to trust and adopt code metrics that show evidence of empirical validation (such as through a technique like \textit{factor analysis}) to accurately measure what they claim to measure. Thus, the results of this study can assist professionals and developers in understanding their code and how code changes impact software quality, providing them with clear quantitative assessments and measures to support their decisions.

For tool developers, we recommend recalibrating the inclusion of metrics and thresholds. Tools must provide clear definitions and detailed implementations of metrics they include including specifying what variant of the metric is being used which is not the current practice. For example, Ardito et al.~\cite{ardito2020} report several software tools do not state what variant of the {\sc lcom} metric they calculate. 
Given the multitude of metrics proposed to measure the same code quality attribute—often yielding conflicting results for the same code—the inclusion of metrics should not be arbitrary nor based solely on their popularity, rather, the evidence of their validity must be the deciding factor for  inclusion. Only metrics that reliably measure their intended construct should be used to establish appropriate thresholds.

\subsection{Limitations}
\label{ssec:limitations}

The recommendations above should be considered in light of several important limitations. Code metrics are not the only way of assessing code quality. Our focus on metrics is not intended as a criticism of any other approach to software quality (e.g. code smells, peer code review, automated testing).  We make no claims about the validity of the nine metrics excluded to avoid an NPD matrix (see \S\ref{ssec:assumptionsfa}). Some of these may be interchangeable with their corresponding, included metrics. We also had to exclude classes for which the tools we studied could not generate data.
The usefulness of a metric in practice has other aspects that can be estimated using additional steps. We only focus on construct validity in this study. We need to determine that a measurement model has construct validity before we can assess whether measuring the target construct is practically useful since usefulness presupposes accurate measurement. 

For high content validity, we need metrics that count \textit{different} things that are all driven by the same construct. For example, for \textit{Size}, the model includes related but different things like {\sc loc} and {\sc nom}, which is desired. In contrast, the three \textit{Cohesion} metrics are very similar--\textit{Cohesion} is measured using {\sc lcom} and its variants; which raises concerns about the content validity of the measure. Using overly similar metrics can create the illusion of convergent validity which is further corroborated by the existence of a heywood case in the CFA. Our current analysis, however, is limited to metrics that Understand, JHawk, and Designite calculate. Future work may involve solutions such as: adding or removing variables from the model, or increasing the sample size. We have established an extremely large sample size compared to the requirement of a factor analysis, so we suggest the addition of variables in the measurement model (removing variables in this case is not recommended since that would violate the three variables per construct rule).

Regarding generalization, we studied one programming language, three metrics tools, and object-oriented, class-level metrics only. We make no claims about other languages, tools, or abstractions. The metrics and tools studied are frequently used in research, but they might not represent the state of the practice.
In terms of representativeness, our sample for the EFA consisted exclusively of the Apache Maven project because applying EFA on a huge dataset is not practical. This is not as serious as it seems because, in the exploratory phase, we only require enough observations (classes) for the factor structure to emerge--the factor structure is a property of the variables (metrics) not the objects of the study (classes).
Our sample for the CFA included multiple Java projects. However, these projects may not fully represent all types of code, as they were open-source projects, that met a predetermined eligibility criteria. Thus, we suggest and encourage replicating our analysis with additional, closed-source and industrial software to examine any potential differences in the results.

Our study also has many strengths. As shown in Tables~\ref{tab:EFAFinal} \&~\ref{tab:loadings}, reliability is high. Verifiability / replicability is also high because we studied a public repository and provide a comprehensive replication package (see \textit{~\nameref{sec:dataavailability}}). We demonstrate novelty by highlighting surprising findings at the beginning of this section. Moreover, while applying the common factor model to code metrics is not entirely new, no prior studies have attempted as comprehensive a model as this one.  

\subsection{Future Work}
\label{ssec:futurework}

This paper reports the first step in a larger research program aiming to improve the validity of code quality measurement. Subsequent steps in this research program include: 
\begin{itemize}
    \item Improving the content validity of the measurement model by including or developing new metrics that reflect existing constructs, but which measure fundamentally different properties from existing metrics. This could include integrating static and dynamic code quality metrics in conjunction with repository mining. 
    \item Extending our analysis to different levels of abstraction (e.g. method, package) using a multi-level model. 
    \item Replicating our analysis on other types of code (e.g. closed-source and industrial software).
    \item Exploring code smell's integration into the model.
    \item Extending the model to additional programming languages. 
\end{itemize}

\section{Conclusion}
\label{sec:conclusion}

To summarize, we used exploratory and confirmatory factor analysis to investigate the construct validity of object-oriented, class-level code metrics generated by three software tools; Designite, JHawk, and Understand. We provide a measurement model with \totalConstructsResult{} constructs--\textit{Cohesion, In-Coupling, Out-Coupling, Size, Sub-Inheritance, and Sup-Inheritance}--each of which is operationalized by at least three metrics. The analysis supports construct validity for \cfatotalMetricsResult{} of the studied metrics, while the evidence for construct validity was lacking for 30 additional metrics. The metrics identified through this study offer empirically supported, quantitative measures to support informed decision making regarding code quality. Using the factor model shown in Table~\ref{tab:loadings} in its present form is scientifically defensible--especially compared to estimating constructs from individual metrics.  However, more research is needed to improve content validity by developing broadened, and more diverse sets of reflective indicators. Taken together, our findings suggest that future software code metrics research should focus on measurement models, instead of interpreting individual metrics as direct measurements in isolation. This study offers the following contributions to the software code quality literature:
\begin{itemize}
    \item Exploratory and confirmatory analysis related to software code quality metrics, providing a comprehensive and novel investigation of their underlying factor structure.
    \item Modeling software quality attributes as latent constructs and applying the common factor model to software code metrics. The resulting measurement model and the replication package provide a foundation for future work in researching construct validity of software metrics.
    \item Highlighting notable findings regarding the relationship between widely cited code quality metrics and the constructs they are intended to measure. 
\end{itemize}

\section*{Data Availability} 
\label{sec:dataavailability}
We provide a comprehensive replication package~\footnote{\url{https://doi.org/10.5281/zenodo.20278447}} which includes the detailed study protocol, scripts, results, and raw code quality metrics data for performing the EFA and CFA.


\bibliography{references}

\end{document}

%% file: Tables/5factors.tex
\begin{table}[!htp]\centering
\caption{Definition of factors at class-level}\label{tab:factorsDefinition}
\scriptsize
\begin{tabular}{lll}\toprule
\textbf{Factor} &\textbf{Description} \\\midrule
\textbf{Size (S)} &The size of the class \\
\textbf{Cohesion (C)} &Degree to which elements in the class belong together \\
\textbf{Inheritance (I)} &Degree of hierarchical class reuse and extension \\
\textbf{In-Coupling (InC)} &Degree to which the class is used by other classes \\
\textbf{Out-Coupling (OutC)} &Degree to which the class is dependant on or uses other classes \\
\bottomrule
\end{tabular}
\end{table}

%% file: Tables/EFAFinal.tex
\begin{table}[h]\centering
\caption{Exploratory Factor Analysis Model}
\label{tab:EFAFinal}
\scriptsize
\begin{tabular}{lcccccccc}\toprule
\textbf{} &\textbf{C} &\textbf{InC} &\textbf{OutC} &\textbf{S} &\textbf{SubI} &\textbf{SupI} &\textbf{h2}  \\\midrule
Cohesion.LCOM.Understand &0.90 & & & & & &0.85 \\
Cohesion.LCOMModified.Understand &0.91 & & & & & &0.82 \\
Cohesion.YALCOM.Designite &0.70 & & & & & &0.5 \\
In-Coupling.CBO.JHawk & &0.85 & & & & &0.96 \\
In-Coupling.FANIN.Designite & &0.98 & & & & &0.93 \\
In-Coupling.FANIN.JHawk & &0.97 & & & & &0.92 \\
Out-Coupling.CBO.Understand & & &0.78 & & & &0.86 \\
Out-Coupling.FANOUT.Designite & & &0.81 & & & &0.89 \\
Out-Coupling.FANOUT.JHawk & & &0.92 & & & &0.96 \\
Size.CountDeclMethodPrivate.Understand & & & &0.76 & & &0.68 \\
Size.CountLine.Understand & & & &0.97 & & &0.98 \\
Size.CountLineBlank.Understand & & & &0.52 & & &0.74 \\
Size.CountLineCodeDecl.Understand & & & &0.81 & & &0.87 \\
Size.CountLineCodeExe.Understand & & & &0.97 & & &0.97 \\
Size.CountLineComment.Understand & & & &0.70 & & &0.49 \\
Size.CountLocalMethodCalls.JHawk & & & &0.67 & & &0.57 \\
Size.CountSemicolon.Understand & & & &0.96 & & &0.94 \\
Size.CountStmt.JHawk & & & &0.98 & & &0.97 \\
Size.CountStmt.Understand & & & &0.99 & & &0.98 \\
Size.CountStmtDecl.Understand & & & &0.86 & & &0.95 \\
Size.CountStmtExe.Understand & & & &1 & & &0.96 \\
Size.LOC.Designite & & & &0.99 & & &0.91 \\
Size.LOC.JHawk & & & &0.98 & & &0.97 \\
Size.LOC.Understand & & & &0.98 & & &0.99 \\
Size.NOM.Designite & & & &0.66 & & &0.79 \\
Size.NOM.Understand & & & &0.68 & & &0.79 \\
Size.UWCS.JHawk & & & &0.59 & & &0.82 \\
Sub-Inheritance.CountSub.JHawk & & & & &0.89 & &0.79 \\
Sub-Inheritance.NC.Designite & & & & &0.62 & &0.49 \\
Sub-Inheritance.NC.Understand & & & & &0.77 & &0.68 \\
Sub-Inheritance.SpecializationRatio.JHawk & & & & &0.86 & &0.76 \\
Sup-Inheritance.CountSup.JHawk & & & & & &0.98 &0.96 \\
Sup-Inheritance.DIT.Understand & & & & & &0.92 &0.88 \\
Sup-Inheritance.ReuseRatio.JHawk & & & & & &0.95 &0.91 \\ \midrule
\textbf{Cronbach's Alpha} &\textbf{0.74} &\textbf{0.95} &\textbf{0.91} &\textbf{0.94}  &\textbf{0.83} &\textbf{0.87} \\
\bottomrule
\end{tabular}
\end{table}

%% file: Tables/loadings.tex
\begin{table}[H]\centering
\sisetup{round-mode=places, round-precision=2}
\caption{Six Construct CFA Result}

\label{tab:loadings}
\scriptsize
\begin{tabular}{@{}lSSSSSS@{}}\toprule
\textbf{} &\textbf{Size} &\textbf{Out-C} &\textbf{In-C} &\textbf{C} &\textbf{Sub-I} &\textbf{Sup-I} \\\midrule
Size.CountDeclMethodPrivate.Understand &0.291 & & & & & \\
Size.CountLine.Understand &0.974 & & & & & \\
Size.CountLineBlank.Understand &0.738 & & & & & \\
Size.CountLineCodeDecl.Understand &0.951 & & & & & \\
Size.CountLineCodeExe.Understand &0.97 & & & & & \\
Size.CountLineComment.Understand &0.364 & & & & & \\
Size.CountLocalMethodCalls.JHawk &0.548 & & & & & \\
Size.CountSemicolon.Understand &0.995 & & & & & \\
Size.CountStmt.JHawk &0.839 & & & & & \\
Size.CountStmt.Understand &0.999 & & & & & \\
Size.CountStmtDecl.Understand &0.968 & & & & & \\
Size.CountStmtExe.Understand &0.997 & & & & & \\
Size.LOC.Designite &0.39 & & & & & \\
Size.LOC.JHawk &0.949 & & & & & \\
Size.LOC.Understand &0.988 & & & & & \\
Size.NOM.Designite &0.363 & & & & & \\
Size.NOM.Understand &0.362 & & & & & \\
Size.UWCS.JHawk &0.542 & & & & & \\
Out-Coupling.CBO.Understand & &0.784 & & & & \\
Out-Coupling.FANOUT.Designite & &0.933 & & & & \\
Out-Coupling.FANOUT.JHawk & &0.954 & & & & \\
In-Coupling.CBO.JHawk & & &0.986 & & & \\
In-Coupling.FANIN.Designite & & &0.805 & & & \\
In-Coupling.FANIN.JHawk & & &0.98 & & & \\
Cohesion.LCOM.Understand & & & &1.002 & & \\
Cohesion.LCOMModified.Understand & & & &0.962 & & \\
Cohesion.YALCOM.Designite & & & &0.5 & & \\
Sub-Inheritance.CountSub.JHawk & & & & &0.988 & \\
Sub-Inheritance.NC.Designite & & & & &0.244 & \\
Sub-Inheritance.NC.Understand & & & & &0.415 & \\
Sub-Inheritance.SpecializationRatio.JHawk & & & & &0.922 & \\
Sup-Inheritance.CountSup.JHawk & & & & & &1.004 \\
Sup-Inheritance.DIT.Understand & & & & & &0.889 \\
Sup-Inheritance.ReuseRatio.JHawk & & & & & &0.884 \\ 
\midrule
\textbf{Average Variance Extracted (AVE)} &0.614695 &0.79842 &0.860207 &0.726483 &0.514497 &0.859931 \\
\midrule
\textbf{Construct Reliability (CR)} &0.961875 &0.921858 &0.948211 &0.880939 &0.772645 &0.948326 \\
\midrule
\textbf{Correlation Matrix*} \\
Size &1 &0.097344 &0.023716 &0.026244 &0.000289 &0.001296 \\
Out-C &0.312 &1 &0.053824 &0.109561 &0.004096 &0.017424 \\
In-C &0.154 &0.232 &1 &0.010609 &0.01 &0.001156 \\
C &0.162 &0.331 &0.103 &1 &0.002116 &0.000049 \\
Sub-I &0.017 &0.064 &0.1 &0.046 &1 &0.000324 \\
Sup-I &0.036 &0.132 &0.034 &0.007 &0.018 &1 \\
\bottomrule

\end{tabular}
\caption*{*Values below the diagonal depict correlations among constructs and values above the diagonal depict squared correlations}
\end{table}